\documentclass[pra,twocolumn,showpacs]{revtex4}
\usepackage{graphicx}
\begin{document}

\bibliographystyle{unsrt}

\title{Ultralow-power nonlinear optics using tapered optical fibers in metastable xenon}
\author{T.B. Pittman}
\author{D.E. Jones}
\author{J.D. Franson}
\affiliation{Physics Department, University of Maryland Baltimore
County, Baltimore, MD 21250}

\begin{abstract}
We demonstrate nanowatt-level saturated absorption using a sub-wavelength diameter tapered optical fiber (TOF) suspended in a gas of metastable xenon atoms. This ultralow-power nonlinearity is enabled by a small optical mode area propagating over a relatively long distance through the Xe gas. The use of inert noble gasses in these kinds of TOF experiments may offer practical advantages over the use of reactive alkali vapors such as rubidium.
\end{abstract}


\maketitle

Sub-wavelength diameter tapered optical fibers (TOF's) and small hollow-core photonic bandgap fibers (PBGF's) enable
low-loss propagation of evanescent and air-guided modes with very small mode areas over very long distances \cite{cregan99,tong04}. The interaction of these highly confined fields with atomic vapors can allow the realization of optical nonlinearities at remarkably low power levels \cite{boydbook}. For example, TOF's surrounded by rubidium vapor, and PBGF's filled with rubidium vapor, have recently been used to demonstrate saturated absorption, two-photon absorption, and a variety of other nonlinear effects at nanowatt and even ``few-photon'' power levels  \cite{ghosh06,spillane08,hendrickson10,saha11,venkataraman11,salit11,venkataraman13}.

Unfortunately, the tendency of Rb to accumulate on silica surfaces \cite{ma09} severely limits the performance of these devices. In the case of TOF's, Rb accumulation causes a drastic loss of transmission \cite{lai13}, while in PBGF's it can limit the penetration depth into the hollow-core to several cm's \cite{venkataraman13}. The observation of these difficulties suggests the use of inert noble gases, rather than reactive Rb vapor, to improve these systems. Here we specifically investigate the use of xenon in TOF experiments. We observe saturated absorption at nanowatt power levels, which indicates the suitability of this system for further ultralow-power nonlinear optics applications.

\begin{figure}[b]
\includegraphics[width=3.15in]{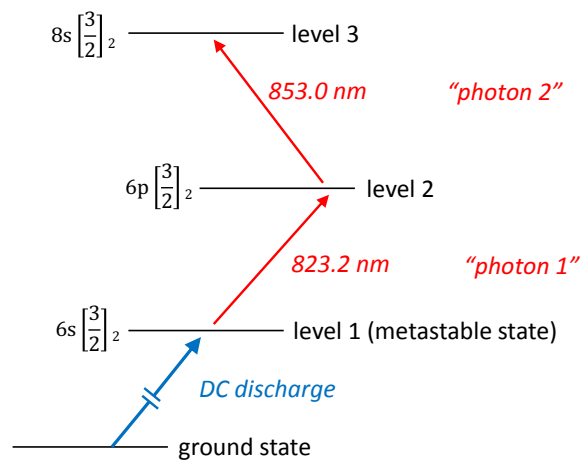}
\caption{(Color online) Diagram of the relevant Xenon energy levels. A DC (or RF) discharge is used to excite the Xe atoms to a long lived metastable state (denoted level 1). A two-photon ladder transition at 823 nm and 853 nm can then used for the experiments of interest.}
\label{fig:energylevels}
\end{figure}

An overview of one particular set of Xe energy levels for these applications is shown in Figure \ref{fig:energylevels}. A weak electric discharge is used to excite Xe to the $6s[3/2]_{2}$ metastable state, which has a long intrinsic lifetime of $\sim$43 s \cite{walhout94}.  This metastable state serves as an effective ``ground state'' for an optical ladder transition at 823 nm and 853 nm that can then be used for the various two-photon nonlinearities. The transition rates of these lines are $3 \times 10^{7}$ s$^{-1}$ and $2 \times 10^{6}$ s$^{-1}$, respectively, which are comparable to those of the well-known $5S_{1/2} \rightarrow 5P_{3/2} \rightarrow 5D_{5/2}$ ladder transition at 780 nm and 776 nm in Rb \cite{NISTatomicdatabase}. In principle, metastable state atomic densities of $10^{13}$ cm$^{-3}$  can be achieved in Xe \cite{uhm08}, which would allow the high optical depths (OD's) desirable for many ultralow-power nonlinear optics applications.

The purpose of this initial work was to perform saturation spectroscopy of the 823 nm transition using a TOF surrounded by a relatively low-density gas of metastable Xe atoms. The ability to saturate this transition at ultralow power levels is an indicator of the overall strength of the atom-field interaction in this system.

\begin{figure}[t]
\includegraphics[width=3.25in]{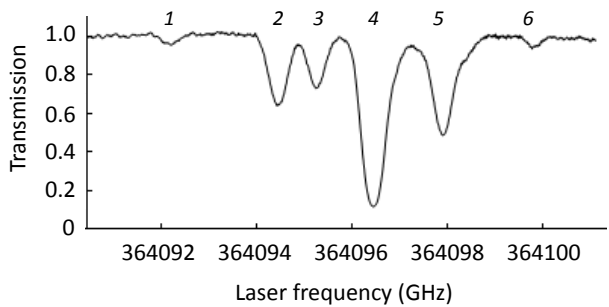}
\caption{Transmission spectrum of metastable Xe at 823 nm obtained using an auxiliary reference discharge cell. The dips labelled 1 through 6 are due to the various stable isotopes of natural Xe. The largest dip (labeled 4) is due to contributions from all of the even isotopes. }
\label{fig:referencecell}
\end{figure}

Figure \ref{fig:referencecell} shows a measured transmission spectrum of the 823 nm line obtained by passing a narrowband tunable diode laser through a conventional Xe discharge tube used as a reference cell. The six primary transmission dips are due to contributions from the 9 stable isotopes of natural Xe \cite{walhout93}.  The largest dip (labeled dip-4) contains contributions from all 7 even isotopes, while the other dips are due to hyperfine splittings of $^{129}$Xe and $^{131}$Xe \cite{xia10}. The reference cell data is used to calibrate our main TOF in Xe spectroscopy system.

\begin{figure}[b]
\includegraphics[width=3.5in]{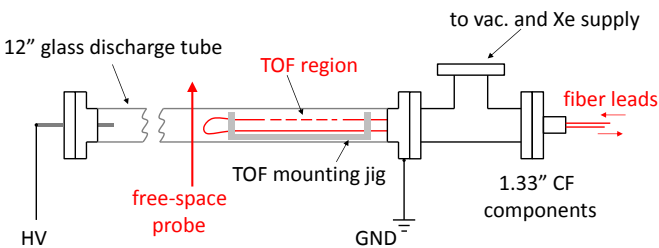}
\caption{(Color online) Design of the DC discharge tube containing a TOF.  A free-space probe beam passing perpendicularly through the discharge tube is used to monitor the density of metastable Xe atoms during the experiments. }
\label{fig:experiment}
\end{figure}

An overview of the main experimental apparatus is shown in Figure \ref{fig:experiment}. TOF's with a minimum waist diameter of 250 nm, and a sub-500 nm diameter over a length of $\sim$1 cm, were pulled from single-mode fiber using the flame-brush technique \cite{birks92}. For the 823 nm wavelength of interest, this TOF size guides an evanescent field with a diameter on the order of 1 $\mu$m over a length of approximately 1 cm \cite{tong04}. This provides an effective ``length to area'' ratio ($L_{eff}/A$) for nonlinear optical effects that is 3 to 4 orders of magnitude larger than comparable free-space focusing to a 1 $\mu$m spot size.

The TOF's were mounted on a thin, 3'' long, cylindrical-shaped Teflon frame using low-outgassing UV-curable epoxy. The TOF mounting jig was then inserted into a 12'' long glass discharge tube system comprised of standard 1.33'' mini-ConFlat (CF) flanges and fittings. An electrical power feedthrough (nickel rod) was used as the discharge tube cathode, and a two-hole fiber feedthrough \cite{abraham98} was used for the input and output fiber leads. Xe pressures of 1 Torr were used for these experiments. A DC power supply (roughly -1 kV) in series with an 8.2 k$\Omega$ ballast resistor was used to maintain the discharge at currents of $\sim$1 mA.

As shown in Figure \ref{fig:experiment}, a free-space probe beam (derived from the primary laser) passed perpendicularly through the discharge tube to monitor the metastable Xe density during the TOF experiments. The probe and TOF output signals were simultaneously recorded as the laser was swept through the Xe resonances. 

\begin{figure}[t]
\includegraphics[width=3.25in]{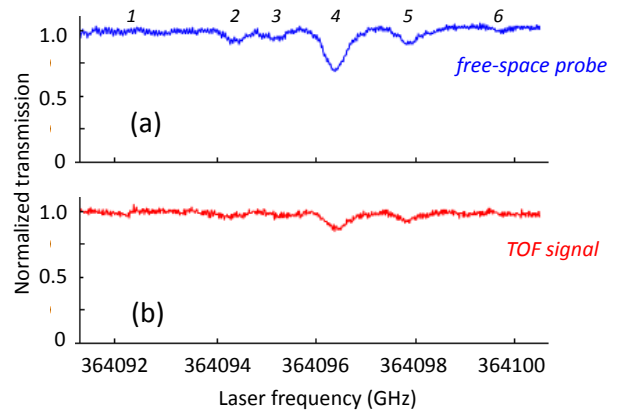}
\caption{(Color online) Transmission spectrum of metastable Xe at 823 nm using the apparatus shown in Figure \protect\ref{fig:experiment}. Plot (a) shows the spectrum of free-space probe beam passing perpendicularly through the discharge tube, while plot (b) shows the spectrum simultaneously obtained by passing a 51 nW signal through the TOF. The absorption dips are labelled 1-6 for comparison with the reference cell data of Figure \ref{fig:referencecell}.}
\label{fig:probeandTOFdips}
\end{figure}

Figure \ref{fig:probeandTOFdips} shows an example result obtained with an estimated power of 51 nW passing through the TOF. The interaction of the TOF evanescent field with the surrounding metastable Xe atoms is apparent by the resonance absorption dips at the same locations as the free-space probe (the measurements were not sensitive enough to observe small shifts or transit-time line broadening effects). We use the largest dip (dip-4) to quantify our measurements of TOF transmission (86\% in this example). When the run was repeated with larger powers in the TOF, the dip-4 transmission was seen to increase towards unity, thereby demonstrating saturation of the system.

\begin{figure}[t]
\includegraphics[width=3.25in]{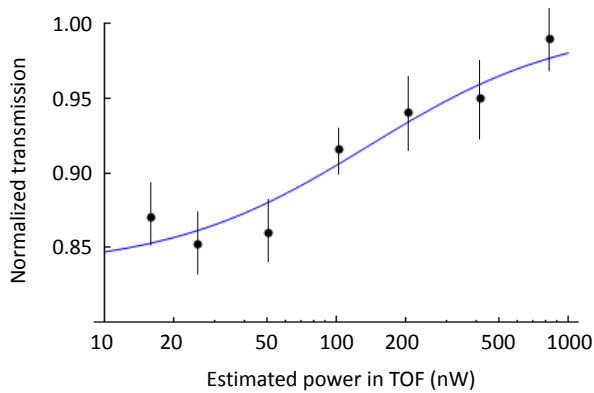}
\caption{(Color online) Measured TOF dip-4 transmission as a function of increasing power passing through the TOF. The data points are best-fit to a simple nonlinear transmission model (blue line) with an ultralow saturation power of 126 nW.}
\label{fig:Psatplot}
\end{figure}

Figure \ref{fig:Psatplot} summarizes this effect. The plot shows the measured TOF dip-4 transmission values for 7 different powers in the range of 10 nW to 1000 nW. The power in the TOF waist region was estimated by assuming the intrinsic TOF loss (74\% for this particular fiber) was uniformly distributed over the entire tapered region, and modifying the measured output power accordingly.   The error bars are due to small uncertainties in maintaining a constant metastable Xe density (as measured by the free-space probe) and noise in the detection system.

The data was fit by a simple transmission model $T = e^{-\alpha_{NL} L}$, with a nonlinear absorption coefficient defined as $\alpha_{NL} = \alpha/(1+P/P_{sat})$ \cite{yarivbook}, and a best-fit saturation power of $P_{sat} = $ 126 nW. The observation of this ultralow saturation power using a TOF in Xe is the main result of this paper. This is comparable to a saturation power of 72 nW that we recently measured with a similar TOF system in Rb vapor \cite{lai13}.

The fitted TOF dip-4 transmission of 84\% (far below saturation) gives an overall system OD of 0.17. Compared with the    free-space probe OD of 0.36 (path length = 12.5 mm), this gives an  $L_{eff} \simeq$ 6 mm for the TOF (here we assume a uniform density of metastable atoms, and neglect any transit-time broadening in the TOF). Given that only $\sim$50\% of the TOF field interacts with the atoms (the remainder is inside the TOF itself)\cite{tong04}, this agrees reasonably well with our SEM-based measurements of a sub-500 nm diameter TOF region over a length roughly 1 cm.

The measured OD's correspond to a relatively low metastable state atomic density in the system. We were able to obtain higher metastable state densities in our system by increasing the DC discharge power, but typically observed a large and irreversible loss of TOF transmission as result. This was presumably due to sputtering of nickel and other contaminants from the electrical feedthrough onto the TOF surface \cite{fujiwara12}. Even at the relatively weak discharge conditions used in our experiments, we typically observed a temporary transmission reduction of roughly 15\% while the discharge was on; most likely due to contaminants passing through the evanescent field. We anticipate that these problems could be minimized by using a better ``right-angle'' DC discharge tube design, or a standard RF discharge system \cite{sukenik02}.

In summary, this initial study suggests the use of metastable Xe gas as a promising alternative to hot Rb vapor for ultralow-power nonlinear optics experiments using TOF's. The two-photon ladder transition of Figure \ref{fig:energylevels} has a number of desirable properties, and the nanowatt-level saturation of the lower transition observed here indicates the suitability of the system for various ``two-beam'' applications \cite{hendrickson10,venkataraman13}. A key next step will be generating much higher metastable state densities (for higher system OD's) without damaging the TOF's. Producing large metastable state Xe densities in small-core PBGF's is also a worthwhile challenge due to the extremely long interaction lengths that may be possible.

This work was supported by DARPA DSO under Grant No. W31P4Q-12-1-0015.



\begin{thebibliography}{50}


\bibitem{cregan99} R.F. Creegan, B.J. Mangan, J.C. Knight, T.A. Birks, P. St. J. Russell, P.J. Roberts, and D.C. Allan, Science {\bf 285} 1537-1539 (1999).

\bibitem{tong04} L. Tong, J. Lou, and E. Mazur, Opt. Exp. {\bf 12}, 1025 (2004).

\bibitem{boydbook} R. Boyd {\em Nonlinear Optics}, Academic Press, New York (1992).

\bibitem{ghosh06} S. Ghosh, A.R. Bhagwat, C.K. Renshaw, S. Goh, A.L. Gaeta, and B.J. Kirby,  Phys. Rev. Lett. {\bf 97} 023603 (2006).

\bibitem{spillane08} S.M. Spillane, G.S. Pati, K. Salit, M. Hall, P. Kumar, R.G. Beausoleil, and M.S. Shariar,  Phys. Rev. Lett. {\bf 100}, 233602 (2008).

\bibitem{hendrickson10} S.M. Hendrickson, M.M. Lai, T.B. Pittman, and J.D. Franson, Phys. Rev. Lett. {\bf 105}, 173602 (2010).

\bibitem{saha11} K. Saha, V. Venkataraman, P. Londero, and A.L. Gaeta, Phys. Rev. A {\bf 83} 033833 (2011).

\bibitem{venkataraman11} V. Venkataraman, K. Saha, P. Londero, and A.L. Gaeta,  Phys. Rev. Lett. {\bf 107}, 193902 (2011).

\bibitem{salit11} K. Salit, M. Salit, S. Krishnamurthy, Y. Wang, P. Kumar,and M.S. Shariar,  Opt. Exp. {\bf 19}, 22874 (2011).

\bibitem{venkataraman13} V. Venkataraman, K. Saha, and A.L. Gaeta, Nature Photonics {\bf 7}, 138-141 (2013).

\bibitem{ma09} J. Ma, A. Kishinevski, Y.Y. Jau, C. Reuter, and W. Happer, Phys. Rev. A {\bf 79}, 4 (2009).

\bibitem{lai13} M.M. Lai, J.D. Franson, and T.B. Pittman, Appl. Opt. {\bf 52} 2595 (2013).

\bibitem{walhout94} M. Walhout, A. Witte, and S.L. Rolston, Phys. Rev. Lett. {\bf 72}, 2843 (1994).

\bibitem{NISTatomicdatabase} A. Kramida {\em et.al.},{\em NIST Atomic Spectra Database}, www.nist.gov/pml/ (2013).

\bibitem{uhm08} H.S. Uhm, P.Y. Oh, and E.H. Choi, Appl. Phys. Lett. {\bf 93}, 211501 (2008).

\bibitem{walhout93} M. Walhout, H.J.L. Megens, A. Witte, and S.L. Rolston, Phys. Rev. A {\bf 48}, R879 (1993).

\bibitem{xia10} T. Xia, S.W. Morgan, Y.-Y. Jau, and W. Happer, Phys. Rev. A {\bf 81} 033419 (2010).

\bibitem{birks92} T.A. Birk and Y.W. Li, J. Lightwave Tech. {\bf 10}, 432-438 (1992).

\bibitem{abraham98} E.R.L. Abraham and E.A. Cornell,  Appl. Opt. {\bf 37}, 1762-1763 (1998).

\bibitem{yarivbook} A. Yariv, {\em Quantum Electronics}, John Wiley and Sons, New York (1989).

\bibitem{fujiwara12} M. Fujiwara, K. Toubara, and S. Takeuchi, Opt. Exp. {\bf 19}, 8596 (2011).

\bibitem{sukenik02} C.I. Sukenik and H.C. Busch, Rev. Sci. Instrum. {\bf 73}, 493 (2002).



\end{thebibliography}
\end{document}